\documentstyle[osa]{revtex}

\begin{document}
\author{V. V. Makhro\thanks{%
e-mail: maxpo@mailexcite.com}}
\address{Bratsk Industrial Institute, Bratsk, Russia, 665709}
\title{Thermally assisted domain walls quantum depinning at the high temperature
range}
\date{}
\maketitle

\begin{abstract}
A theoretical and numerical investigations of the quantum tunneling of the
domain walls in ferromagnets and weak ferromagnets was performed taking into
account the interaction between walls and thermal excitations of a crystal.
The numerical method for calculations of the probability of a thermally
stimulated quantum depinning as the function of temperature at the wide
temperature range has been evolved.
\end{abstract}

Macroscopic quantum effects in magnetism are currently of great interest.
Such phenomena are important in tests of quantum mechanics \cite{leggett}.
In particular, magnetic domain walls tunneling seems to be one of most
appropriate subject for investigations in this field. Together with detailed
theoretical and experimental researches of domain walls in ferromagnets\cite
{eg}\cite{chud}\cite{stamp}\cite{gior}\cite{CIS} , recently the similar
phenomena were described in a weak ferromagnets\cite{zvez}.

For the description of the domain walls dynamics it is convenient to apply
the model, in which the wall is considered as a quasiparticle with some
effective mass $m$. Such quasiparticle transferred via the crystal the
change of a magnetic moments orientation. In the movement through a crystal
the quasiparticle can be trapped by magnetic pinning center - as provided
for example, by an impurity raising the anisotropy energy locally. The
domain wall then can overcome this energy barrier in a following ways:
either due to absorption of an external fields energy or due to thermal
activation, and at last via tunneling.

Tunneling and thermal activation are usually considered as competing
processes; thereat one think tunneling can be observable only at extremely
low temperatures about 0.001 K - 1 K, whereas at higher temperatures
tunneling is suppressed by thermal activation. However, it is not always
true. The purpose of this paper is the theoretical and numerical
investigations of the situations in which to some extent both discussed
phenomena can cooperate. How that one comes to think of it, due to
interaction with thermal excitations of a crystal and absorption of their
energy, the wall ''raised'' in front of barrier. In this case effective
height of barrier will decrease and, accordingly tunneling rate will
increase. Further we spend detailed discussion of this mechanism for the
Bloch walls and walls in weak ferromagnets.

\section{ Bloch walls thermally activated tunneling}

\subsection{ Model and equations of motion}

Let us consider in the beginning 180-degrees Bloch wall in an uniaxial
crystal. The form of the rest wall has given by well-known Landau-Lifshitz%
\cite{lanZov} exact solution

\begin{equation}
\sin \theta =\tanh (x/\sqrt{A/K_1})\text{ ,}
\end{equation}

where $x$ is directed along easy axis, $A$ is constant of exchange and $K_1$
is constant of uniaxial anisotropy. For a travelling wall an additional
''kinetic''energy $K$ arises. It may be represented in quadratic on velocity 
$v$ form $K=v^2m_d/2,$ where $m_d=E_g/8\pi A\gamma ^2\sin ^2\theta $ is
so-called D\"oring mass\cite{der} or, rather, surface density of the mass,
and $E_g=\sqrt{AK_1}$ is surface energy density of a wall. In this terms
equation of motion took on Newton's form

\[
m_d\frac{d^2X}{dt^2}=2I_sH\text{ ,} 
\]

where $X$ is coordinate of center of a wall. Let us consider interaction
between wall and defect introducing the potential energy $U(X)$ ; then last
equation takes the form

\begin{equation}
m_d\frac{d^2X}{dt^2}=2I_sH-\frac{dU(X)}{dX}\text{.}
\end{equation}

In this paper we consider point-like repulsive impurity. In this case the
potential energy of the wall will be change in manner $U(\theta )=U_0\cos
^2\theta $, where $U_0$ is maximal value of potential energy attainable when
impurity lies in a center of the wall. If one take into consideration eq.
(1) it is easy to obtain explicit function $U(X)$:

\begin{equation}
U(X)=U_0(1-\tanh ^2(X/\sqrt{A/K_1}))=U_0\cosh ^{-2}(X/a)\text{,}
\end{equation}

where $a=\sqrt{A/K_1}$ - characteristic width of both domain wall and
potential $U(X)$.

When eq. (2) was written, it was supposed that structure of moving wall does
not vary and definition of $m$ contains energy of the rest wall. However,
Walker\cite{walk} has shown that this assumption is justified only for
slowly travelling walls. As it follows from exact solutions of the equation
of motion for the Bloch wall, structure of a wall (in particular, its width
and mass) strongly depends on velocity of movement. When walls velocity
tends to some critical value $c$, derivative of walls energy reduce to
infinity, its mass tends to infinity and width tends to zero. Mostly, in
usual ferromagnets the Walker limiting velocity $c$ has value about several
kilometers per a second, and in this case $mc^2\gg U_0$ . In this reason one
may consider that condition $v^2/c^2\ll 1$ in ferromagnets usually is
satisfied, therefore in current section we shall be limited by a case of
small velocities.

In a context of the discussed problem, the following physical situation will
be interested us. Let the domain wall with kinetic energy $K$ arrived at a
potential barrier with the height $U_0$ , which simulates its interaction
with a defect of the crystal. If $K<U_0$ the segment of wall in immediate
proximity to barrier will be trapped in a metastable minimum. It is
acceptably to consider such segment as an isolated quasiparticle with the
effective mass determined as an integral over the area of defect\cite{stamp}.

As stated above, there are three ways for the wall to overcoming barrier:
with the help of an external field, due to thermal activation and via
tunneling. Let us consider very weak field $H\ll \frac 1{I_s}\left| \frac{%
\partial U}{\partial X}\right| $. Such field can't to disengage the wall,
but it will create asymmetry for a displacement of a wall in front of
barrier and behind of barrier. For the invariability of the walls structure,
the condition $U_0\ll mc^2$ must be satisfied. On the other hand, height of
barrier should not be too small inasmuch as thermal activation will become
the prevail and tunneling will be suppressed. Together, both these
restrictions give a rather narrow corridor for reasonable parameters.
Nevertheless, values which we shall accept are usual for ferromagnets,
therefore it is possible to use them for calculations. Let us suppose height
of barrier $U_0=10^{-14}$ erg, quasiparticles mass $m=10^{-26}$ gm , width
of potential $a=10^{-6}$ cm and area of defect $S=10^{-13}cm^2.$

\subsection{Calculations and results}

By virtue of the preset assumption on weakness of an external field, it is
possible to believe that metastable minimums width is lot greater than
barrier's width, whereas its depth is considerably smaller than the barrier
height. In such case the quasiparticles in front of barrier has a
quasi-continuous spectrum and using of Maxwellian distribution for the
analysis of a problem is acceptable.

Let us suppose a quasiparticle in a thermal equilibrium with a crystal. We
consider an ensemble containing $N$ such particles. We shall perform
computations in according to the next computational scheme. Interval of
energy from 0 up to $U_{\max }\footnote{$U_{\max }$ was adjusted so as there
will be neglected number of particles outside interval for each given value
of a temperature.}$ was divided into equal subintervals $\delta w.$ The
number of particles for each subinterval one can found from the expression

\begin{equation}
N_w=\int_{w-\delta w/2}^{w+\delta w/2}\frac{2N}{\sqrt{\pi }}(k_BT)^{-1.5}%
\sqrt{w}\exp (-\frac w{k_BT})dw.
\end{equation}

In the next step we assign to all particles within the limits of given
subinterval an identical average value $w$. The validity of this device was
checked numerically, i.e. number of dissections ${\it N}$ adjusted so as to
guarantee the stability of computing scheme as a whole.

Further, the barrier penetrability $D$ was calculated for each subinterval
by well-known formulae\cite{LanUch}

\begin{equation}
\begin{array}{c}
D= \frac{\sinh ^2(\pi ka)}{\sinh ^2(\pi ka)+\cosh ^2(\frac \pi 2\sqrt{\frac{%
32\pi ^2U_0ma^2}{h^2}-1})}\text{ when }\frac{32\pi ^2mU_0a^2}{h^2}<1\text{
and } \\ 
D=\frac{\sinh ^2(\pi ka)}{\sinh ^2(\pi ka)+\cos ^2(\frac \pi 2\sqrt{1-\frac{%
32\pi ^2U_0ma^2}{h^2}})}\text{ when }\frac{32\pi ^2mU_0a^2}{h^2}>1
\end{array}
\end{equation}

where $k=\frac{\sqrt{2mw}}\hbar $. We emphasize that eq. (5) is the exact
solution of Shr\"odinger equation for the potential (3).

Product $D\times N_w$ gives $N_{w0}$ or a number of particles from given
subinterval which overcomes the barrier. The total sum of all $N_{w0}$ gives 
$N_0$ - number of all particles in ensemble transmitted the barrier. Then $%
F=N_0/N$ will be effective barrier penetrability. Let us note that magnitude
of $F$ is determined not only by tunneling but by over-barrier reflection
too. Really, even in case when energy of particle large then $U_0$ , $D$ may
be less then 1.

The probability of thermal activation $G$ one can easily found if to
calculate fraction of particles with energy above $U_0$%
\begin{equation}
\begin{array}{c}
G=\int_{U_0}^\infty \frac 2{\sqrt{\pi }}(k_BT)^{-1.5}\sqrt{w}\exp (-\frac w{%
k_BT})dw. \\ 
\end{array}
\end{equation}

In numerical calculations on upper bound we of course use substitution $%
\infty \rightarrow U_{\max }$ with precautions described above.

The dependencies both of effective barrier penetrability $F$ and probability
of thermal activation $G$ on the temperature are plotted in fig. 1 and 2.
Fig. 1 represented influence of discussed mechanism in wide temperature
range from 0 to 300 K. In fig. 2 for most clearness plotted same results in
narrow region - from 0 to 20 K, where exist greatest difference between $F$
and $G$ . We hope that these results demonstrated severity of the tunneling
exposure to the depinning processes.

\section{Thermostimulated domain walls tunneling in a weak ferromagnets.
Accounting of the quasi-relativism.}

As stated above the thermal stimulation of the tunneling for Bloch walls can
be realized only under rigid restrictions, especially for parameter $U_0.$
For high barriers the bordering $U_0\ll mc^2$ may be broken and structure of
the wall will be varied; the quasi-relativistic phenomena arise from this
reason. The accounting of such peculiarities will be demonstrated by example
of walls in a weak ferromagnet. Theory, which successfully described
high-energy dynamics in such materials, was evolved in Ref.\cite{zvezSlab}.
The weak ferromagnets has very suitable properties for comparison theory
with experimental data. Walls in weak ferromagnets, as a rule, has a mass by
one or two order of magnitude smaller then in ferromagnets, and its width
essentially smaller too. Both this factors leads to increasing of the
tunneling rate. Let us also denote, that very pure samples with low defects
concentration are available now, this factor gives good reproducibility of
measurements results.

\subsection{Model}

Let us consider, for example, a weak ferromagnet of a therbium orthoferrits
type within the two-lattice approximation using the ferro- and
antiferromagnetics vectors ${\bf m}$ and ${\bf l}$, respectively. Its
thermodynamical potential will be\cite{zv12}

\[
\begin{array}{c}
\Phi _0( {\bf l,m})=Jm^2+A(\nabla {\bf l})^2-{\bf mH}%
+d_1m_xl_z-d_3m_zl_x+K_{ac}l_z^2+K_{ab}l_x^2\text{, } \\ 
\end{array}
\]

where $J$ and $A$ are, respectively, constants of uniform and non-uniform
exchange, $K_{ac}$ and $K_{ab}$ - constants of anisotropy, ${\bf H}$ - total
external field acting on the wall, $d_1$ and $d_3$ are Dzyaloshinsky
exchanges constants. After minimization on ${\bf m}$ one can obtain\cite
{zv12}\cite{zv13}

\[
\begin{array}{c}
\Phi _0=A(\nabla {\bf l})^2-\frac{\chi _{\bot }^2}2(H^2-({\bf HL}%
)^2)-M_z^0H_zl_x-M_x^0H_xl_z+K_{ac}l_z^2-K_{ab}l_{x^{}}^2\text{,} \\ 
\end{array}
\]

where $\chi _{\bot }$ is transverse susceptibility, $M_x^0$ and $M_z^0$ are
the values of magnetization in phases $\Gamma _4({\bf l\Vert x})$ and $%
\Gamma _2({\bf l\Vert z})$ respectively. Without loss of generality, we can
consider $ac$-type walls only. In spherical coordinates corresponding
Lagrange density will be\cite{bar}

${\cal L}=\frac{\chi _{\bot }}{2\gamma ^2}(\frac{\partial {\bf l}}{\partial t%
})^2-\frac{\chi _{\bot }}\gamma {\bf H[l,}\frac{\partial {\bf l}}{\partial t}%
{\bf ]}-\Phi _0$

The appropriate Lagrange function per unit area of the domain wall has
essentially quasi-relativistic form\cite{zvezSlab}

\begin{equation}
L=-m^{*}c^2\sqrt{1-v^2/c^{*2}}-U(x)
\end{equation}

where $m^{*}c^{*2}=4\sqrt{AK\text{ }}$ and $c^{*}=\gamma \sqrt{A/\chi _{\bot
}}$ is spin-wave velocity; $U(x)$ - potential of wall-defect interaction.
Associated with eq. (7) Hamiltonian then will be

\begin{equation}
H_g(p,x)=c^{*}\sqrt{p^2+m^{*2}c^{*2}}+U(x),
\end{equation}

where $p=\frac{mv}{\sqrt{1-v^2/c^{*2}}}$ - canonical impulse. The kinetic
energy then will be $K=pc^{*}.$

The quasi-relativistic type of Hamiltonian (8) carried to essential
peculiarities in wall dynamics. It is important for us that energy
distribution for walls (quasiparticles) will appreciably changed, and
effective barrier penetrability $F$ and probability of thermal activated
depinning $G$ may changed too. Therefore it is necessary to define energy
distribution for a quasi-relativistic particles. For this purpose we shall
use the Gibbs canonical distribution

\[
dz_p=a^0\exp (-K(p)/k_BT)dp, 
\]

where $a^0$ one can found by using normality condition $\int_{-\infty
}^\infty dz_p=1.$ As since $K(p)=cp$ then

\begin{equation}
z(w)=\frac 1{2(k_BT)^3}w^2\exp (-\frac w{k_BT}).
\end{equation}

Obviously, maximum of this distribution in comparison with the same for
classical distribution will shifted into high-energy region. Therefore
fraction of ''vigorous'' particles will increase and correspondingly will
changed $F$ and $G.$

\subsection{Results of calculations}

From results of numerical calculations submitted below it will be visible
that the walls in weak ferromagnets are really more suitable subjects for
the study of the tunneling in the high temperature range. Let us designate
the main parameters. Effective mass of a wall let will be $%
m=10^{-13}gm/cm^2. $ We shall take both width of an energy barrier of the
defect and width of a wall $\Delta $ of the same order $10^{-6}cm.$ Both
area of the defect and area of the tunneling segment of the wall will be $%
S=10^{-13}cm^2$, hence quasiparticle mass will be $m^{*}=10^{-26}gm.$ The
height of barrier one can find from the value of coercive force\cite{zvez} : 
$U_0=10^{-13}erg.$ On the face of it such values seems little suitable for
the tunneling. After formal substitution this parameters in eq. (5) one can
found negligible range for $D $ - about $10^{-80}$ or smaller. But account
of the quasi-relativism essentially changed the physical situation.

Let us consider this situation more detailed. Now, as before, we shall carry
out modeling of the barrier by function (3). The sorting particles of an
ensemble by the energy we shall execute in according with algorithm
presented in Sec. 1. But taking into account eq. (9) we find now for $N_w$

\begin{equation}
N_w=\int_{w-\delta w/2}^{w+\delta w/2}\frac{Nw^2}{2(k_BT)^3}\exp (-\frac w{%
k_BT})dw.
\end{equation}

Accordingly, the probability of thermal activated depinning will be

\[
G=\int_{U_0}^\infty \frac{w^2}{2(k_BT)^3}\exp (-w/k_BT)dw 
\]

The calculations of $D$ for a wall in a weak ferromagnet were done
numerically with using of the technique offered in Ref.\cite{zvez}. The
final computational results for $F$ and $G$ are plotted in figure 3. It is
visible that with accepted parameters the difference between probability of
quantum thermal-stimulated depinning $F$ and thermal activated depinning $G$
more distinct than in usual ferromagnets. In particular, at 50 K $F$ two
times as large $G,$ and even at room temperature difference between $F$ and $%
G$ amounts to 0.05. Thus, account of quantum effects in mechanism of the
depinning is actually topical.

In Sec. 1 we have considered Bloch walls tunneling in usual ferromagnet. The
results obtained there are valid for low-energy walls only. At high energy
walls dynamics also becomes quasi-relativistic and for its description Walker%
\cite{walk} technique is necessary. The structure of Walker solution
corresponds formally to Hamiltonian (9), therefore one should think
reasonable to propagate results of Sec. 2 on the Bloch walls. However,
nature of maximal velocity in this case is other. In ferromagnet, when walls
velocity tends to maximal (not limiting) value, the form of the wall can
essentially changed. In this case it is necessary to take into account
additional energy, connected with a leakage fields, because of propagation
of the results of the given section into Bloch walls required an additional
analysis.

\section{Admissibility of the experimental testing}

Domain walls depinning via tunneling was investigated usually at very low
temperatures (see, for example, Review\cite{stamp}). Unfortunately, we have
no data on a tunnel depinning at high temperatures. The data concerning to
the basic parameters of a problem, such as form, width and height of the
energy barrier was obtained indirectly and at present time is not
sufficiently reliable. This fact made difficult the comparison theory with
experimental data. Therefore the special importance has experiments, where
the depinning of solitary walls is investigated. Problems, connected with a
statistical nature of a depinning, will be in this case eliminated. In this
respect it is very interesting experimental technique used in ref.\cite{gior}%
, where a solitary wall tunneling in a superthin wire was investigated. In
our opinion, expansion this technique on high temperature region may be very
useful for testing physical mechanism described in present paper. Testing of
the presented results may be carried out also with using the magnetic noise
technique. At any rate, this is in urgent need of search for the departures
from classical temperature dependence for any physical quantities associated
with walls depinning.

Author is grateful to A K Zvezdin and V V Dobrovitski for helpful discussion.

\end{document}